\documentclass[aps,pra,groupaddress,twocolumn,floatfix,reprint]{revtex4-1}
\usepackage{units}
\usepackage{amsmath}
\usepackage{amssymb}
\usepackage{graphicx}
\usepackage{bm}
\usepackage{multirow,color,relsize,ulem,microtype}

\newcommand{\be}{\begin{equation}}
\newcommand{\ee}{\end{equation}}

\newcommand{\im}[1]{\text{Im}[#1]}

\newcommand{\cc}[1]{{\color{black}#1}}

\begin{document}
\title{Parity-time symmetry in a flat band system}

\author{Li Ge}
\email{li.ge@csi.cuny.edu}
\affiliation{Department of Engineering Science and Physics, College of Staten Island, CUNY, Staten Island, New York 10314, USA}
\affiliation{The Graduate Center, CUNY, New York, New York 10016, USA}
\date{\today}

\begin{abstract}
In this paper we introduce Parity-Time ($\cal PT$) symmetric perturbation to a one-dimensional Lieb lattice, which is otherwise $\cal P$-symmetric and has a flat band. In the flat band there are a multitude of degenerate dark states, and the degeneracy $N$ increases with the system size. \cc{We show that the degeneracy in the flat band is completely lifted due to the non-Hermitian perturbation in general, but it is partially maintained with the half-gain-half-loss perturbation and its ``V" variant that we consider.} With these perturbations, we show that both randomly positioned states and pinned states at the symmetry plane in the flat band can undergo thresholdless $\cal PT$ breaking. They are distinguished by their different rates of acquiring non-Hermicity as the $\cal PT$-symmetric perturbation grows, which are insensitive to the system size. Using a degenerate perturbation theory, we derive analytically the rate for the pinned states, whose spatial profiles are also insensitive to the system size. Finally, we find that the presence of \cc{weak disorder} has a strong effect on modes in the dispersive bands but not on those in the flat band. The latter respond in completely different ways to the growing $\cal PT$-symmetric perturbation, depending on whether they are randomly positioned or pinned.
\end{abstract}

\maketitle

\section{Introduction}

Systems that exhibit flat bands have attracted considerable interest in the past few years, including optical \cite{Manninen1,Manninen2} and photonic lattices \cite{Rechtsman,Vicencio,Mukherjee,Biondi}, graphene \cite{Kane,Guinea},  superconductors \cite{Simon,Kohler1,Kohler2,Imada}, fractional quantum Hall systems \cite{Tang,Neupert,Sarma} and exciton-polariton condensates \cite{Jacqmin,Florent}. One interesting consequence of a flat band is the different scaling properties of its localization length when compared with a dispersive band \cite{Flach,Leykam,PRB15}, due to a multitude of degenerate states in the flat band. This degeneracy has another important implication on Parity-Time ($\cal PT$) symmetry breaking \cite{Bender1,Bender2,Bender3,El-Ganainy_OL06,Moiseyev,Kottos,Musslimani_prl08,Makris_prl08,Longhi,CPALaser,conservation,Robin,RC,Microwave,Regensburger,Kottos2,Peng,Feng,Hodaei}:
it was found recently that the degeneracy of the underlying Hermitian spectrum, before any $\cal T$-breaking perturbations are introduced, determines whether thresholdless $\cal PT$ symmetry breaking is possible \cite{PRX14}. \cc{Therefore, it is interesting to probe the interplay of the large degeneracy in a flat band system and $\cal PT$ symmetry breaking. In particular, we would like to know whether the degeneracy is completely lifted due to a $\cal PT$-symmetric perturbation, and whether $\cal PT$ symmetry breaking depends on the evenness (oddness) of the degeneracy and its magnitude that grows with the system size. In addition, because a flat band makes the underlying Hermitian system more susceptible to disorder, it is equally important to understand the role of disorder on $\cal PT$ symmetry breaking in a flat band system.}

\begin{figure}[t]
\begin{center}
\includegraphics[clip,width=\linewidth]{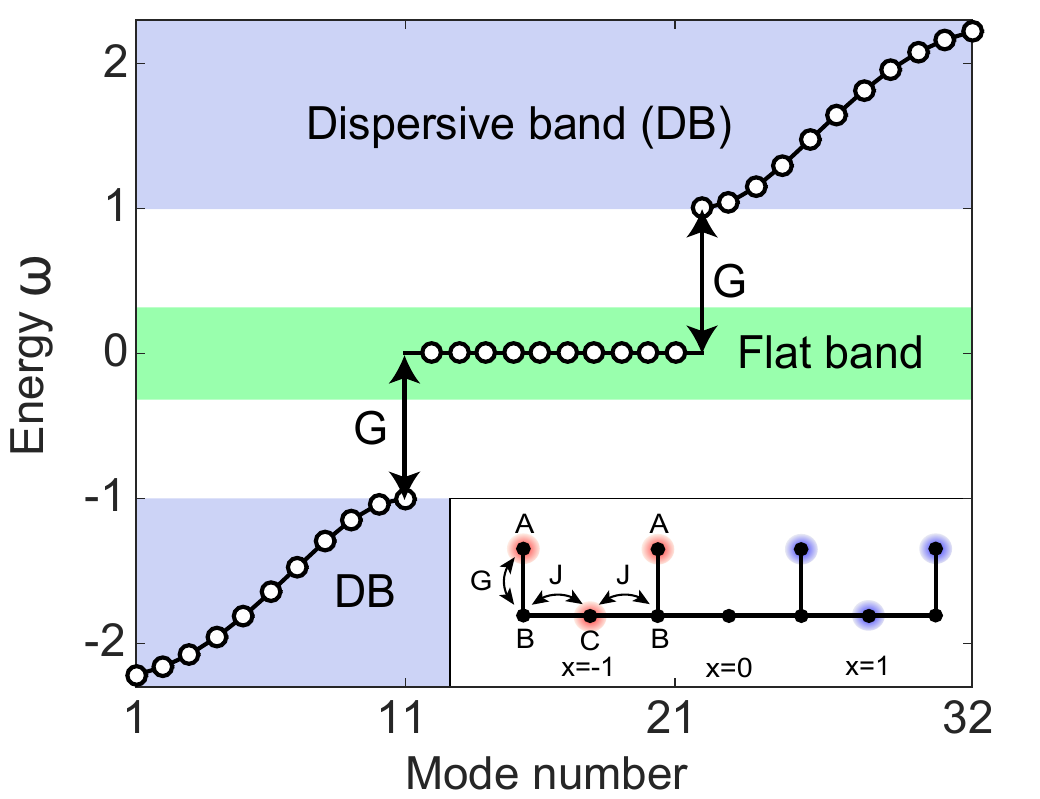}
\caption{(Color online) Inset: A quasi-1D Lieb lattice with $3$ overlapping ``$A$--$B$--$C$--$B$--$A$" sublattices. Two dark states are shown schematically by circular clouds, which also serve as an example of a $\cal PT$-symmetric ``V" perturbation to be discussed in Section II.
Main figure: Energy eigenvalues in a Lieb lattice with 10-fold degeneracy in the flat band. There is one more mode in each of the two dispersive bands. \cc{In all figures the energy is scaled by the coupling $G(=J)$, where the onsite energies are all shifted to zero ($\omega_A=\omega_B=\omega_C=0$).}} \label{fig:schematic}
\end{center}
\end{figure}

Using a quasi-one-dimensional (quasi-1D) Lieb lattice (see Fig.~\ref{fig:schematic}) and \cc{a half-gain-half-loss perturbation (including its variant, the ``V" configuration to be introduced below)}, we show in this paper that two different scenarios of thresholdless $\cal PT$ symmetry breaking can take place, depending on whether the degeneracy $N$ in the flat band is even or odd. When $N$ is odd, all but one flat band modes enter the $\cal PT$-broken phase at the infinitesimal strength of the $\cal PT$-symmetric perturbation. They form two $(N-1)/2$ degenerate branches, each confined strictly to either the gain half or the loss half of the lattice, and their non-Hermicity equals the strength of the $\cal PT$-symmetric perturbation, which we denote by $\tau$. \cc{The remaining flat band mode experiences an exceptional point of order 3 \cite{Graefe}, via the coupling with two dispersive band modes.} When $N$ is even, all flat band modes experience thresholdless $\cal PT$ breaking, but now they exhibit four branches. Two branches are ($N/2-1$) degenerate and have the same properties as those in the $N$-odd case. There is only one mode in each of the remaining two branches, and they form a $\cal PT$-symmetric doublet, pinned at the symmetry plane ($x=0$) with exponential tails in both the gain and the loss halves. Surprisingly, we find that this localization is not due to the half-gain-half-loss nature of the $\cal PT$-symmetric perturbation as previously found \cite{PRA11}, but rather a result of the point defect at $x=0$ of this perturbation to satisfy the $\cal PT$ symmetry. Therefore, this spatial profile is insensitive to both $\tau$ and the system size when $N$ is large, and it can be reproduced in the underlying Hermitian system at $\tau=0$.
Using a degenerate perturbation theory, we derive this localization length and the rate these doublet states acquire non-Hermicity.

Finally, we show that the presence of \cc{weak disorder} has a strong effect on modes in the dispersive bands but not on those in the flat band. The finite $\cal PT$-transition thresholds of the former are smoothed out, while the thresholdless $\cal PT$ breaking of the latter is largely preserved. In addition, the flat band modes respond in completely different ways to increasing $\cal PT$-symmetric perturbation when there is disorder. The $\cal PT$-symmetric perturbation has little effect on them if they are already localized to either half of the lattice. Otherwise the perturbation forces them to pick a side, unless they are the doublet states, which evolve from Anderson localized states \cite{Anderson} to those pinned by the point defect at $x=0$. We note that different from Refs.~\cite{Makris_prl08,Szameit, Zhen}, the system we consider here has a flat band \textit{before} the $\cal PT$-symmetric perturbation is introduced, instead of being the result of $\cal PT$-symmetry breaking.

\section{Thresholdless $\cal PT$ breaking in a Lieb lattice}
\label{sec:II}
In two-dimensional systems of a finite size, the maximum degeneracy due to a point group is 2. In a flat band, however, the degeneracy can be arbitrarily large, because it increases with the system size. Take a quasi-1D Lieb lattice for example (see Fig.~\ref{fig:schematic}), where an upright $A$ site is decorated to every other site of a 1D chain (i.e., the $B$ sites). A flat band exists when the onsite energy of $A$ and $C$ sites are the same, $\omega_A =\omega_C\equiv\omega_0$ ($\hbar=1$), and $\omega_0$ gives the flat band energy. It is gapped from the two dispersive bands by $G$, the real-valued coupling between two neighboring $A$ and $B$ sites. For simplicity we take $\omega_B=\omega_0$ as well. Because $\omega_0$ shifts all eigenvalues by the same amount and has no effect on the eigenstates, we take it to be zero.
The degeneracy $N$ of the flat band in this Lieb lattice equals the number of existing dark states, which we will denote by $\vec{V}(x)$ due to their geometry (see the inset in Fig.~\ref{fig:schematic}). Each dark state has a nonvanishing amplitude only at a $C$ site and two nearest $A$ sites:
\be
\vec{V}(x) = [J,0,-G,0,J]^T,\label{eq:darkstate}
\ee
where the five elements are the amplitudes of the wave function on a ``$A$--$B$--$C$--$B$--$A$" sublattice, to which we will refer as an $U$ sublattice below. The $B$ sites are black in $\vec{V}(x)$ because the tunneling probabilities from the neighboring $A$ and $C$ sites cancel each other. The argument $x$ in $\vec{V}(x)$ is the position of the central $C$ site of an $U$ sublattice in the units of the lattice constant. It takes integer or half integer values, depending on whether $N$ is odd or even, and the center of the lattice is placed at $x=0$.
We note that the number of (overlapping) $U$ sublattices equals the number of dark states, and each end of the lattice is terminated on $A$ and $B$ sites.
The superscript ``$T$" in Eq.~(\ref{eq:darkstate}) denotes the matrix transpose, and $J$ is the real-valued coupling between two neighboring $B$ and $C$ sites. Below we drop the vector symbol of $\vec{V}(x)$ without causing ambiguity.

The tight-binding Hamiltonian of the Lieb lattice can be written as
\begin{align}
\bm{H}_0 = &\sum_{j,Z}\omega_Z|Z_j\rangle\langle Z_j| + \sum_j\left[\,G|A_j\rangle \langle B_j| + h.c.\,\right] + \nonumber \\
&\sum_j \left[\,J(\,|B_j\rangle \langle C_j| + |C_j\rangle\langle B_{j+1}|\,) + h.c.\,\right],
\end{align}
where $j$ runs through all unit cells, $Z=A,B,C$, and $h.c.$ denotes Hermitian conjugate of the other terms in the square brackets. We then introduce the non-Hermitian perturbation $i\tau \bm{H}_1$, which is $\cal PT$-symmetric about $x=0$. $\tau$ is the overall strength of the perturbation, and $\bm{H}_1$ is a diagonal matrix with positive (gain), negative (loss), and zero (no gain or loss) elements. To satisfy the $\cal PT$-symmetry, no gain or loss is introduced on the lattice sites right at the symmetry plane at $x=0$, which include one $A$ and one $B$ site when $N$ is even and a single $C$ site when $N$ is odd.

\begin{figure}[b]
\begin{center}
\includegraphics[clip,width=\linewidth]{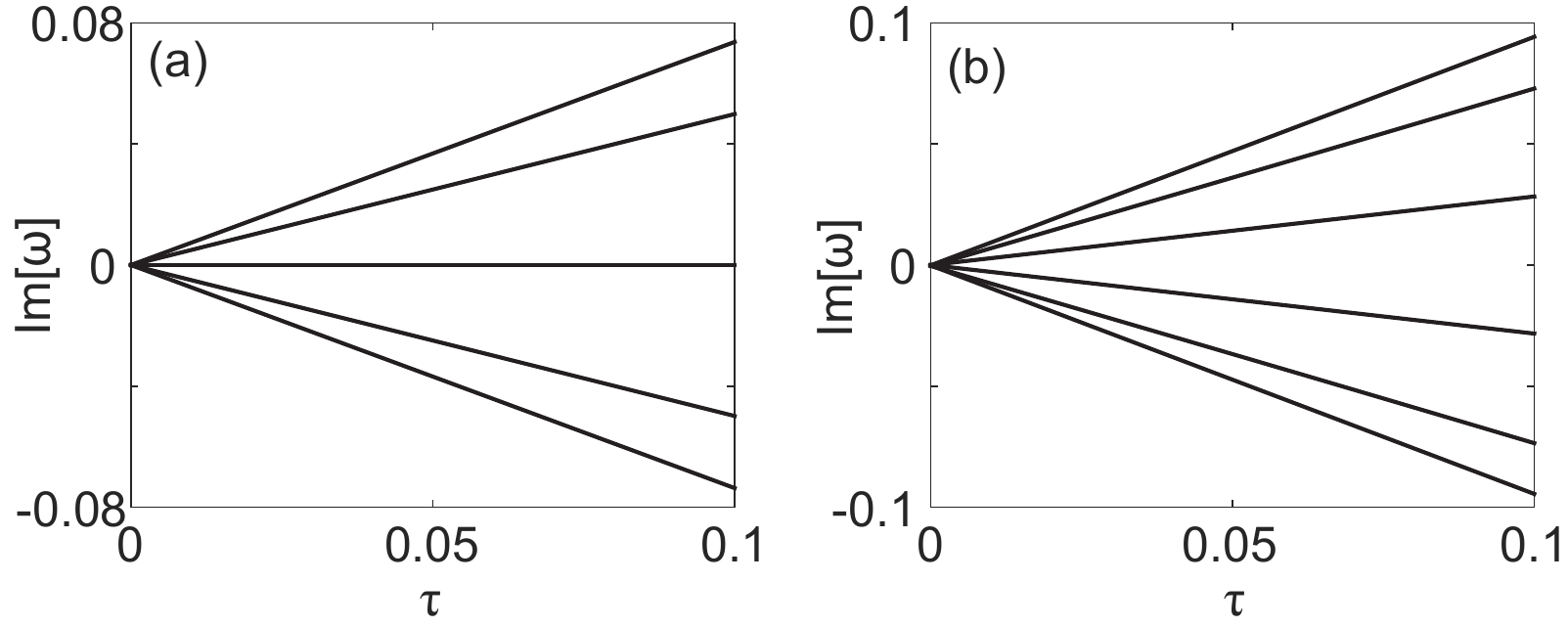}
\caption{\cc{Thresholdless $\cal PT$ breaking in a Lieb lattice that completely lifts the degeneracy in the flat band with a general $\cal PT$-symmetric configuration for (a) $N=5$ and (b) $N=6$.} }\label{fig:breaking_nonV}
\end{center}
\end{figure}

\cc{The degeneracy of the flat band modes is completely lifted by $\bm{H}_1$ in general as they undergo thresholdless $\cal PT$ symmetry breaking, independent of whether $N$ is even or odd (see Fig.~\ref{fig:breaking_nonV}).
The simplest $\cal PT$-symmetric perturbation, with uniform gain on one half of the whole lattice and the same amount of loss on the other half, in fact partially maintains the degeneracy of the flat band. A variant of this half-gain-half-loss configuration with the same property is the ``V" configuration (see Fig.~\ref{fig:schematic}; inset), which differs by having no gain or loss on all the $B$ sites (where the flat band modes are dark).}

\begin{figure}[t]
\begin{center}
\includegraphics[clip,width=\linewidth]{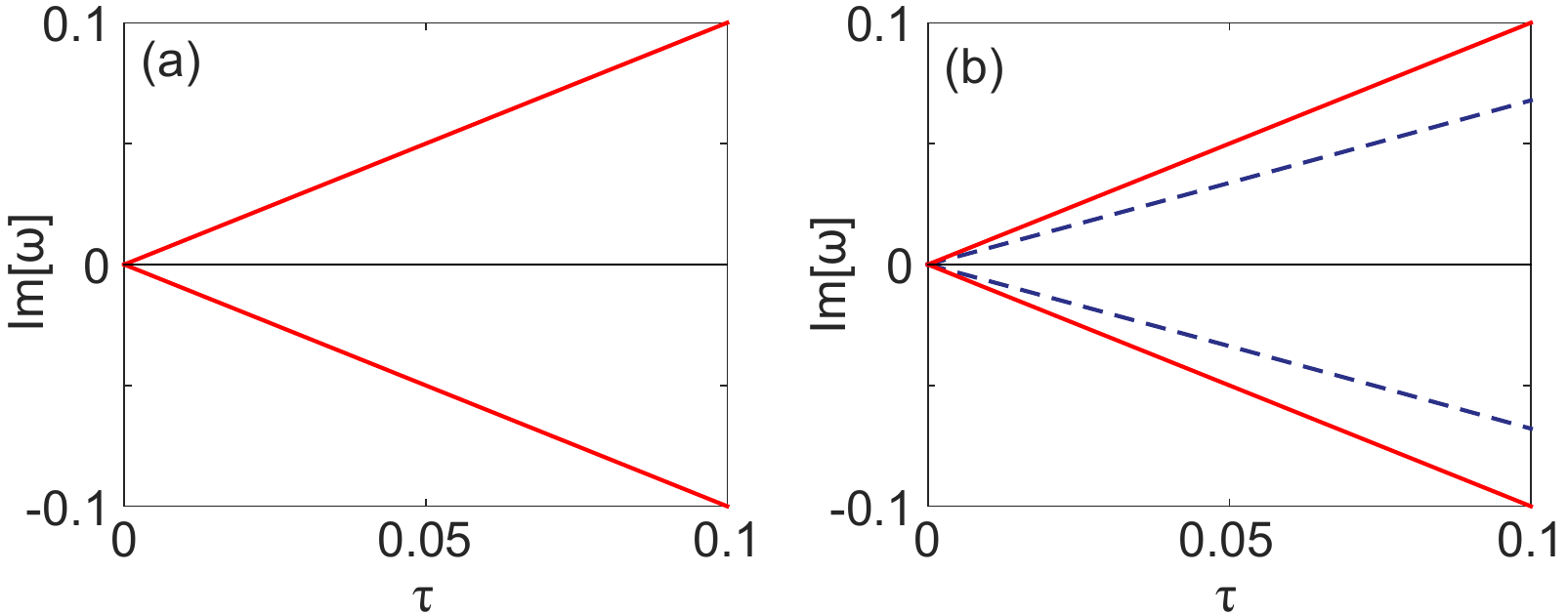}
\caption{(Color online) Thresholdless $\cal PT$ breaking in a Lieb lattice with the ``V" configuration for (a) $N=5$ and (b) $N=6$. (a) 4 flat band modes enter the $\cal PT$-broken phase at $\tau=0$ (thick solid lines). The other flat band mode and all 12 dispersive band modes remain in the symmetric phase in the range of $\tau$ shown (thin solid line). (b) All 6 flat band modes enter the $\cal PT$-broken phase at $\tau=0$, 4 with $\kappa_\pm=\pm\tau$ (thick solid lines) and 2 with $0<|\kappa_\pm|<\tau$ (dashed lines). All 14 dispersive band modes remain in the symmetric phase in the range of $\tau$ shown (thin solid line).}\label{fig:breaking}
\end{center}
\end{figure}

\begin{figure}[t]
\begin{center}
\includegraphics[clip,width=\linewidth]{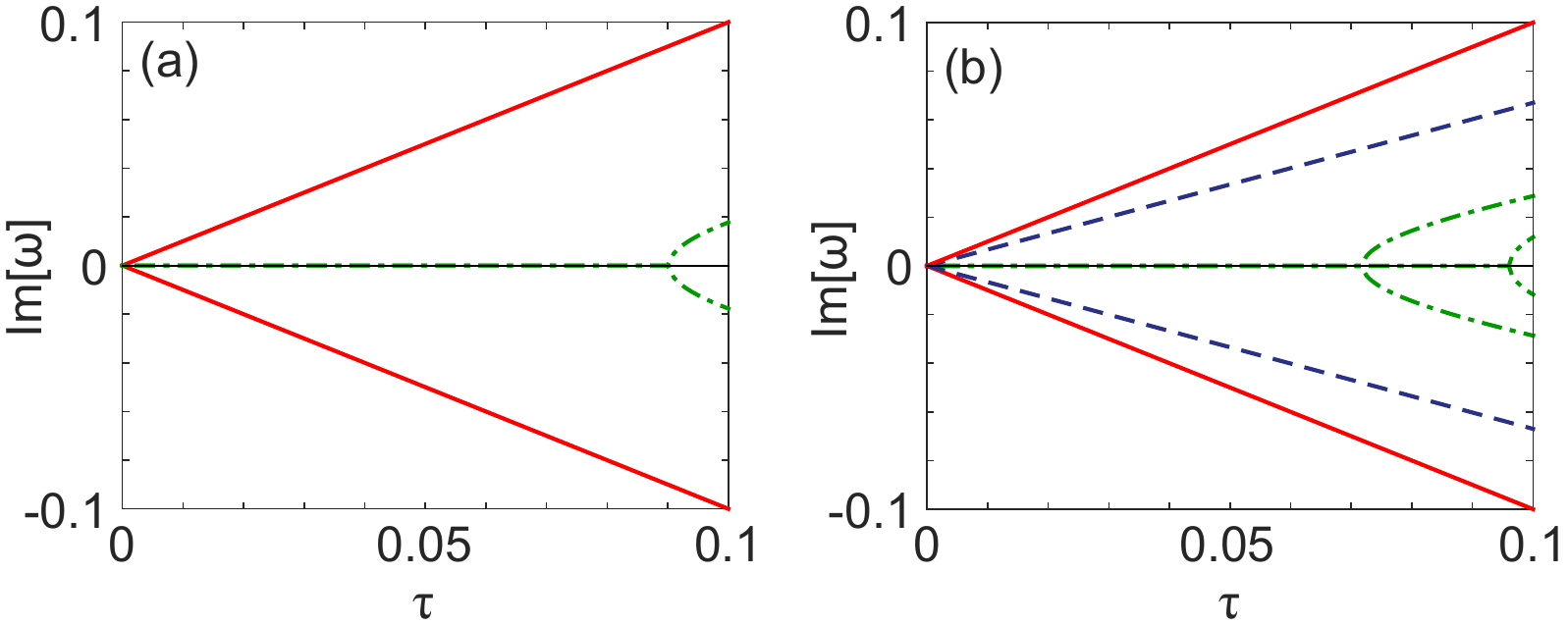}
\caption{(Color online) Same as Fig.~\ref{fig:breaking} but with (a) $N=7$ and (b) $N=8$. (a) 4 and (b) 8 dispersive band modes enter the $\cal PT$-broken phase before $\tau=0.1$ (dash-dotted lines). Note that the features of the flat band modes (solid and dashed lines) are the same as in Fig.~\ref{fig:breaking}. The horizontal thin lines represent modes in the $\cal PT$-symmetric phase.}\label{fig:breaking2}
\end{center}
\end{figure}

\cc{To see how the half-gain-half-loss configuration and the ``V" configuration partially maintain the degeneracy of the flat band, we first note that a pair of $V(x),V(-x)$ modes are $\cal PT$-symmetric partners, i.e., $V(x)={\cal PT}V(-x)$, because they are real-valued wave functions (see Eq.~(\ref{eq:darkstate})). In addition, they are also eigenfunctions of $\bm{H}_1$ for these two configurations, with opposite eigenvalues ($\pm1$) when they do not overlap with the $A$ and $C$ site at $x=0$. Therefore, they undergo thresholdless $\cal PT$ symmetry breaking simultaneously as $\tau$ increases from 0, and the imaginary parts of the corresponding energy eigenvalues are nothing but $\kappa_\pm\equiv\im{\tilde{\omega}_\pm}=\pm \tau$ (see Fig.~\ref{fig:breaking}), independent of the system size and the couplings $G,J$. In other words, their non-Hermiticy (given by $|\kappa_\pm|$) is the same as the strength of the $\cal PT$-symmetric perturbation (i.e., $\tau$). There are $(N-1)/2$ such pairs of flat band modes when $N$ is odd (e.g., 2 and 3 pairs for the $N=5,7$ cases shown in Figs.~\ref{fig:breaking}(a) and \ref{fig:breaking2}(a)) and $(N/2-1)$ such pairs when $N$ is even (e.g., 2 and 3 pairs for the $N=6,8$ cases shown in Fig.~\ref{fig:breaking}(b) and \ref{fig:breaking2}(b)),
which all behave in the same way. As a result, each of the two $\kappa_\pm=\pm \tau$ branches has $(N-1)/2$ ($N$-odd) or $N/2-1$ ($N$ even) degeneracy in the $\cal PT$-broken phase. Such a large number of degenerate states in the $\cal PT$-broken phase has not been reported before.}

\begin{figure}[t]
\begin{center}
\includegraphics[clip,width=\linewidth]{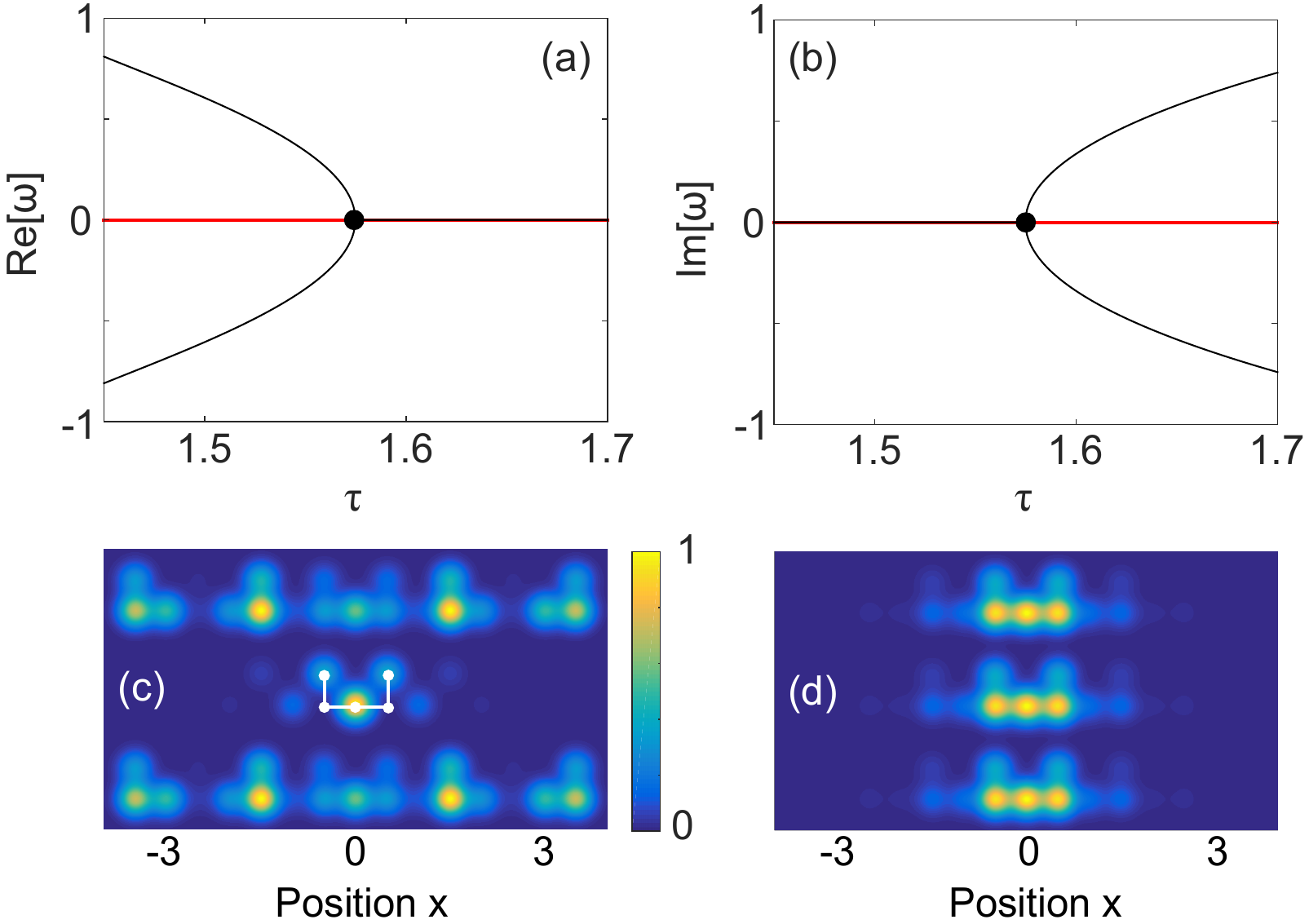}
\caption{\cc{(Color online) EP$_3$ in a Lieb lattice with the ``V" configuration and $N=7$. The real and imaginary parts of the energy eigenvalues of two dispersive band modes and the solitary dark state are shown in (a) and (b) respectively, with the EP$_3$ marked by the black dot. False color plots of the absolute value of their wave functions at (c) $\tau=0$ and (d) 1.58 are shown. The white $U$ sublattice in (c) indicates different lattice sites.}} \label{fig:EP3}
\end{center}
\end{figure}

For the single dark state $V(x=0)$ that is left alone in the $N$-odd case, it is no longer an eigenstate of $\bm{H}_1$ since there is no perturbation on the $C$ site at $x=0$ in order to respect the $\cal PT$ symmetry. Hence it needs to couple to modes in the dispersive bands to break the $\cal PT$ symmetry, which then leads to a finite threshold in terms of $\tau$. \cc{In fact at this $\cal PT$-symmetry breaking threshold lies an exceptional point of order 3 (EP$_3$) \cite{Graefe}, where the energy eigenvalues and wave functions of three modes coalesce, including two dispersive band modes and the solitary dark state (see Fig.~\ref{fig:EP3}). Nevertheless, a hybridized mode remains in the $\cal PT$-symmetric phase after the EP$_3$, as if the solitary dark state remained in the $\cal PT$-symmetric phase. We note that this EP$_3$ does not occur if the onsite energy $\omega_B$ is detuned from $\omega_A$ and $\omega_C$ or with the half-gain-half-loss configuration.}

The two dark states $V(x=-1/2)$ and $V(x=1/2)$ in the $N$-even case are not eigenstates of $\bm{H}_1$ either, due to the absence of perturbation on the $A$ site at $x=0$ in the ``V" configuration (\cc{and the $B$ site at $x=0$ as well in the half-gain-half-loss configuration}) in order to respect the $\cal PT$ symmetry. We will refer to these two states as the doublet states. Although they do not belong to the ($N/2-1$) degenerate manifolds of the two $\kappa_\pm=\pm\tau$ branches, they still experience a thresholdless $\cal PT$ breaking, but their $|\kappa_\pm|$ is smaller than $\tau$ (see Fig.~\ref{fig:breaking}(b)). Interestingly, we find that the slope of these $|\kappa_\pm|$ is also insensitive to the system size $N$, when $N>2$. More specifically, this slope is about $0.708,0.675,0.671,0.670$ when $N=2,4,6,8$, measured by $\kappa'_+(\tau=0.1)\approx\left.\frac{\kappa_+(\tau)}{\tau}\right|_{\tau=0.1}$ with the ``V" configuration (see Fig.~\ref{fig:breaking}(b)), \cc{which is the configuration we will focus on below.}

\begin{figure}[t]
\begin{center}
\includegraphics[clip,width=\linewidth]{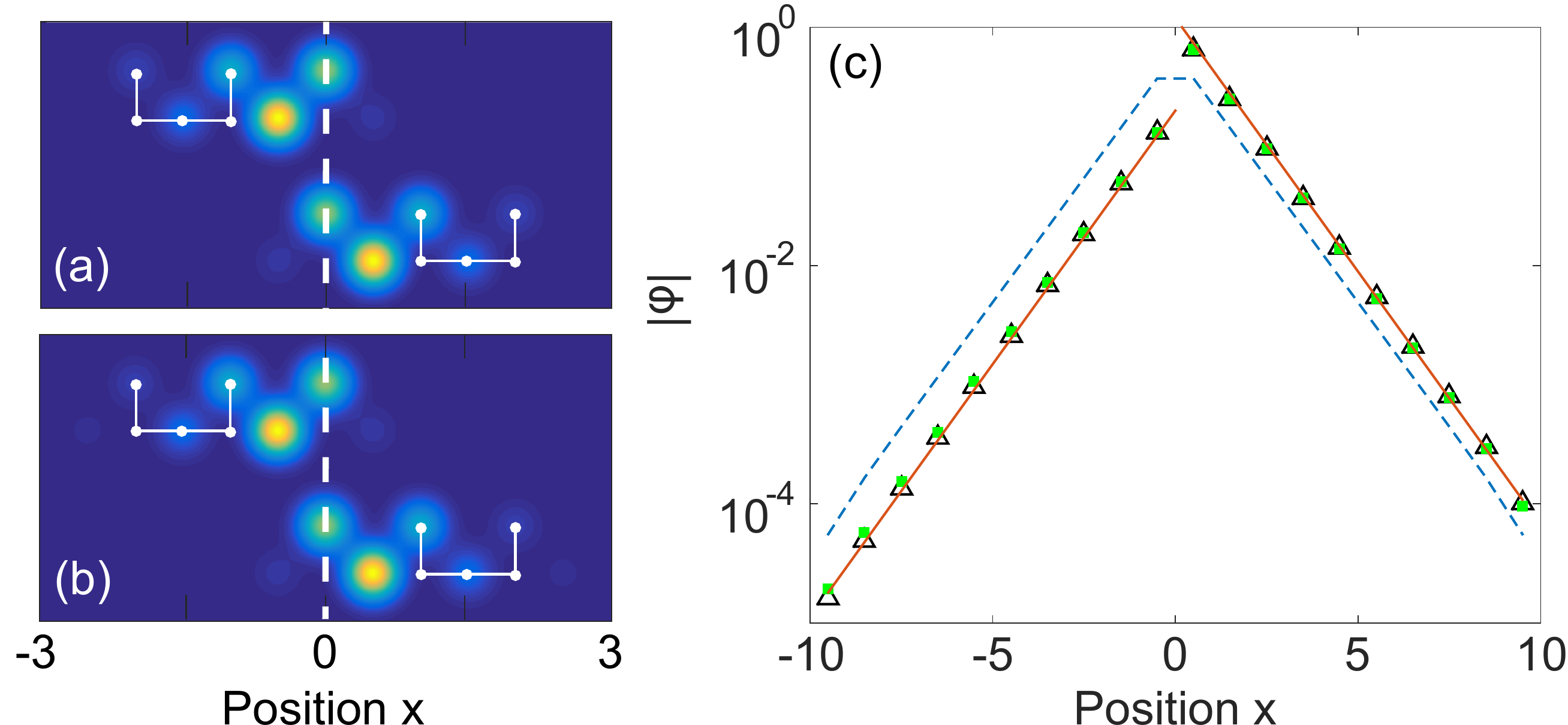}
\caption{(Color online) Doublet states in the $\cal PT$-broken phase with the ``V" configuration. (a) False color plot of the absolute value of their wave functions for $N=6$. The vertical dashed line marks $x=0$, and the color scale is the same as in Figs.~\ref{fig:EP3}(c) and (d). (b) Same as (a) but for $N=20$. (c) Exponential tails of the amplifying doublet state shown in (b) (solid lines). Only the absolute values of the wave function on the $C$ sites are shown. Circular dots and open triangles are calculated at $\tau=0.002$ and $0.2$, respectively. The solid lines are proportional to $\exp(-|x|/1.03)$ shown for comparison. The dashed line shows the defect state at $\tau=0$ by increasing $\omega_A(x=0)$ to 0.1.} \label{fig:wf}
\end{center}
\end{figure}

To understand this property, we resort to two approaches, a qualitative one based on directly visualizing the spatial profiles of the doublets states, and a quantitative one based on a degenerate perturbation theory. As Fig.~\ref{fig:wf}(a) shows, the doublet states are pinned near $x=0$ and have tails in both the gain and the loss halves of the lattice. We find that these spatial profiles are also insensitive to the system size when $N$ is large enough ($>2$; see Figs.~\ref{fig:wf}(a) and (b)), which explains the same property of their $\kappa_\pm$.
In fact, we find the tails of their spatial profiles decay exponentially away from $x=0$ (see Fig.~\ref{fig:wf}(c)). One might expect that these exponential tails are a result of gain and loss, as it has been shown that the wave function in the $\cal PT$-broken phase tends to peak at the gain and loss interface \cite{PRA11}. However, we find that the exponential tails are largely independent of the $\cal PT$-symmetric perturbation strength, either. As Fig.~\ref{fig:wf}(c) shows, the tails are captured well by an exponent of $-|x|/1.03$, and  there is little change in the wave function when $\tau$ changes from $0.002$ to $0.2$. Instead, we find that this localization length of $\xi=1.03$ is due to the point defect of the ``V" configuration at $x=0$, which has neither gain or loss on the $A$ site at $x=0$ in order to respect the $\cal PT$ symmetry as mentioned. If we introduce a point defect at $x=0$ in the Hermitian system at $\tau=0$, for example, by increasing $\omega_a$ here to 0.1, we recover the same localization length (see the dashed line in Fig.~\ref{fig:wf}(c)).

Before we derive the perturbation theory for the doublet states, we briefly discuss the modes in the dispersive bands. They have high $\cal PT$-transition thresholds when $N$ is small. This observation can be understood in the following way: their eigenvalues are distributed in $|\omega|\in[\omega_0+G, \omega_0+\sqrt{G^2+4J^2}]$ (see Fig.~\ref{fig:schematic}) and are well separated from each other when $N$ is small; their thresholds are proportional to these spacings, in the simplest case of two-mode coupling \cite{PRX14}. As $N$ increases, their energy spacings reduce and so do their $\cal PT$-transition thresholds. For example, no dispersive band modes enter the $\cal PT$-broken phase in Fig.~\ref{fig:breaking}, but their lowest threshold reduces to $\tau\approx0.090$ and $0.072$ for $N=7$ and $8$, respectively (see Fig.~\ref{fig:breaking2}). Due to the symmetry of the dispersive bands about $\omega_0=0$ (see Fig.~\ref{fig:schematic}), when one pair of dispersive band modes enter the $\cal PT$-broken phase at some finite value of $\tau$, there is always another pair that do the same at exactly the same $\tau$ but on the opposite side of the flat band energy $\omega_0$; their $\kappa_\pm$ are nevertheless the same. \cc{Different from the EP$_3$ scenario depicted in Fig.~\ref{fig:EP3} in which two dispersive band modes are also involved, here the real parts of their energy eigenvalues are different from $\omega_0$ in the $\cal PT$-broken phase (not shown).}

\section{Degenerate perturbation theory}
\label{sec:perturbation}
Now we turn to the perturbation theory for a more quantitative understanding of the doublet states, and hence the $N$ considered in this section is even. We start by diagonalizing the perturbation $\bm{H}_1$ in the $N$-fold flat band subspace of $\bm{H}_0$, which is required in the degenerate perturbation theory \cite{Shanker}. In the $N=2$ case, we define the basis in this subspace as $\phi_1^{(2)} = V(-1/2)+aV(1/2)$ and $\phi_\cc{2}^{(2)} = bV(-1/2)+V(1/2)$, where we can take $a,b$ to be real numbers, because $\bm{H}_1$ (without the factor $i\tau$) is Hermitian. The superscript enclosed in parentheses indicates the value of $N$. The above requirement then means $\langle \phi_1^{(2)}|\bm{H}_1|\phi_2^{(2)}\rangle=0$, and an additional constraint is $\langle \phi_1^{(2)}|\phi_2^{(2)}\rangle=0$. We note that this requirement is different from demanding that $\phi_1^{(2)},\phi_2^{(2)}$ are the eigenstates of $\bm{H}_1$; the latter is sufficient but unnecessary.

Since the flat band states are dark on $B$ sites, we can project the wave functions and $\bm{H}_1$ to the Hilbert space of $A$ and $C$ sites, leading to $V(-1/2)=[J,-G,J,0,0]^T$, $V(1/2)=[0,0,J,-G,J]^T$, and $\bm{H}_1 = \text{diag}(1,1,0,-1,-1)$ when $N=2$, where ``$\text{diag}$" stands for a diagonal matrix. Using $\langle \phi_1^{(2)}|\bm{H}_1|\phi_2^{(2)}\rangle=0$ we find $a=b$, and $\langle \phi_1^{(2)}|\phi_2^{(2)}\rangle=0$ leads to
\be
a = \pm\sqrt{\left(2+\frac{G^2}{J^2}\right)^2-1} - \left(2+\frac{G^2}{J^2}\right).
\ee
Note that the two values of $a$ are reciprocal to each other, which means that if one leads to $\phi_1^{(2)}$, the other one gives $\phi_2^{(2)}$. By neglecting the coupling to the dispersive band modes, we then derive
\be
\left|\kappa^{(2)}_\pm(\tau)\right| = \left|\frac{\langle\phi_1^{(2)}|\bm{H}_1|\phi_1^{(2)}\rangle}{\langle\phi_1^{(2)}|\phi_1^{(2)}\rangle}\right|\tau= \left|\frac{a^2-1}{2a}\right|\frac{J^2}{3J^2+G^2}\tau \label{eq:kappa2}
\ee
for the doublet states, which gives a slope of $0.707$ when $G=J=1$. Using $\phi_2^{(2)}$ in the expression above leads to the same result. Previously we have mentioned that the numerically obtained slope in this case is $0.708$, calculated using $\kappa'_+(\tau=0.1)\approx\left.\frac{\kappa_+(\tau)}{\tau}\right|_{\tau=0.1}$. If we reduce the perturbation strength at which the slope is calculated, for example, to $\tau=0.02$, we find $\left.\frac{\kappa_+(\tau)}{\tau}\right|_{\tau=0.02}=0.707$, which agrees nicely with the analytical result given by Eq.~(\ref{eq:kappa2}). Since the slope of $|\kappa_\pm|$ given by Eq.~(\ref{eq:kappa2}) is independent of $\tau$, we know that any change to the slope must be a result of coupling to the dispersive bands, which is neglected in the derivation of Eq.~(\ref{eq:kappa2}). Nevertheless, the small difference between $\frac{\kappa_+(\tau)}{\tau}$ at $\tau=0.02$ and 0.1 indicates that such coupling is weak.

For $N>2$, the construction of the first $N-2$ basis functions in the $N$-fold flat band subspace is straightforward: any dark states that do not overlap with $x=0$ are eigenfunctions of $\bm{H}_1$, as we have mentioned. These $N-2$ dark states just need to be linearly superposed properly such that they form $N-2$ orthogonal basis functions, and there is more than one way to achieve it.
For the remaining two doublet states, we use mathematical induction to find their approximate forms. Assuming that we have solved the $(N-2)$ case and found the correct doublet states $\phi_1^{(N-2)},\phi_2^{(N-2)}$, we then approximate the correct doublet states in the $N$ case by
\begin{align}
\phi_1^{(N)} &\approx \phi_1^{(N-2)} + cV(-N/2) + dV(N/2),\\
\phi_2^{(N)} &\approx \phi_2^{(N-2)} + dV(-N/2) + cV(N/2),
\end{align}
where $c,d$ are two real numbers. The basic assumption is that the central part of $\phi_1^{(N)},\phi_2^{(N)}$ near $x=0$ is insensitive to $N$, as we have seen in Figs.~\ref{fig:wf}(a) and (b). It is easy to check that $\langle\phi_1^{(N)}|\bm{H}_1|\phi_2^{(N)}\rangle=0$, using $\langle\phi_1^{(N-2)}|\bm{H}_1|\phi_2^{(N-2)}\rangle=0$, and by requiring $\langle\phi_1^{(N)}|\phi_2^{(N)}\rangle=0$, we find
\be
cd(2J^2+G^2) + 2J(c\delta^{(N-2)}+d\epsilon^{(N-2)})=0,
\ee
where $\epsilon^{(N-2)},\delta^{(N-2)}$ are the first (last) and last (first) elements of $\phi_1^{(N-2)}$ ($\phi_2^{(N-2)}$). An additional constraint imposed by the degenerate perturbation theory is that $\phi_1^{(N)},\phi_2^{(N)}$ need to be orthogonal to \cc{the first $N-2$ basis functions in the $N$-fold flat band subspace. Let us pick, for example, one such basis function simply as $\phi_3^{(N)}=V(-N/2)$ (which is an eigenfunction of $\bm{H}_1$), and we immediately find $d=-\delta^{(N-2)}/3$ from $\langle\phi_3^{(N)}|\phi_1^{(N)}\rangle=0$ and $c=-\epsilon^{(N-2)}/3$ from $\langle\phi_3^{(N)}|\phi_2^{(N)}\rangle=0$ when $G=J=1$. We note that we arrive at the values of $c$ and $d$ with other choices of $\phi_3^{(N)}$, such as $V(N/2)$ or $V(-N/2)-V(N/2)$.}
Once $c$ and $d$ are found, we can immediately construct the leftmost and rightmost elements of $\phi_1^{(N)}$ and $\phi_2^{(N)}$ to derive the localization length for the doublet states. For example, the first four elements of $\phi_1^{(N)}$ are $cJ,-cG,cJ+\epsilon^{(N-2)},-\epsilon^{(N-2)}G/J$. The localization length $\xi$ can then be defined as
\be
\xi^{-1} = \ln\left|\frac{\epsilon^{(N-2)}}{cJ}\right|,\label{eq:xi}
\ee
using the values of the wave function on the first $C$ site (i.e., $-cG$) and the second $C$ site (i.e., $-\epsilon^{(N-2)}G/J$) from the left. Equation ~(\ref{eq:xi}) gives $\xi = \ln3 \approx 1.10$, which agrees reasonably well with the result of numerical fitting of the exponential tails shown in Fig.~\ref{fig:wf}(c), i.e., $\xi=1.03$. \cc{In the Appendix we give another way to estimate this localization length.}

For the slope $\kappa'_\pm(\tau)$ of the doublet states, a recursive relation can then be formulated. Assuming $\phi_1^{(N)}$ is amplifying, we find
\begin{align}
{\kappa'}^{(N)}_+ &= \frac{\langle\phi_1^{(N)}|\bm{H}_1|\phi_1^{(N)}\rangle}{\langle\phi_1^{(N)}|\phi_1^{(N)}\rangle}\nonumber \\
&=\frac{3{\kappa'}^{(N-2)}_+ -\frac{{\delta^{(N-2)}}^2-{\epsilon^{(N-2)}}^2}{\cc{\langle\phi_1^{(N-2)}|\phi_1^{(N-2)}\rangle}}}{3-\frac{{\delta^{(N-2)}}^2+{\epsilon^{(N-2)}}^2}{\cc{\langle\phi_1^{(N-2)}|\phi_1^{(N-2)}\rangle}}}.
\end{align}
It is clear that ${\kappa'}^{(N)}_+$ (and similarly ${\kappa'}^{(N)}_-$) becomes insensitive to $N$ as $N$ becomes large, because $\epsilon^{(N-2)},\delta^{(N-2)}$ are at the very ends of the exponentially decaying tails of $\phi_1^{(N-2)}$, i.e., $|\delta^{(N-2)}|^2,|\epsilon^{(N-2)}|^2\ll{\langle\phi_1^{(N-2)}|\phi_1^{(N-2)}\rangle}$,  leading to a converging series ${\kappa'}^{(N)}_+\approx{\kappa'}^{(N-2)}_+$.
Using ${\kappa'}^{(2)}_+=0.707$ given by Eq.~(\ref{eq:kappa2}), we successively find ${\kappa'}^{(4)}_+=0.673$, ${\kappa'}^{(6)}_+=0.669$, ${\kappa'}^{(8)}_+=0.669$, which agree well with the previously mentioned numerical values.

\section{Effect of disorder}

\begin{figure}[t]
\begin{center}
\includegraphics[clip,width=\linewidth]{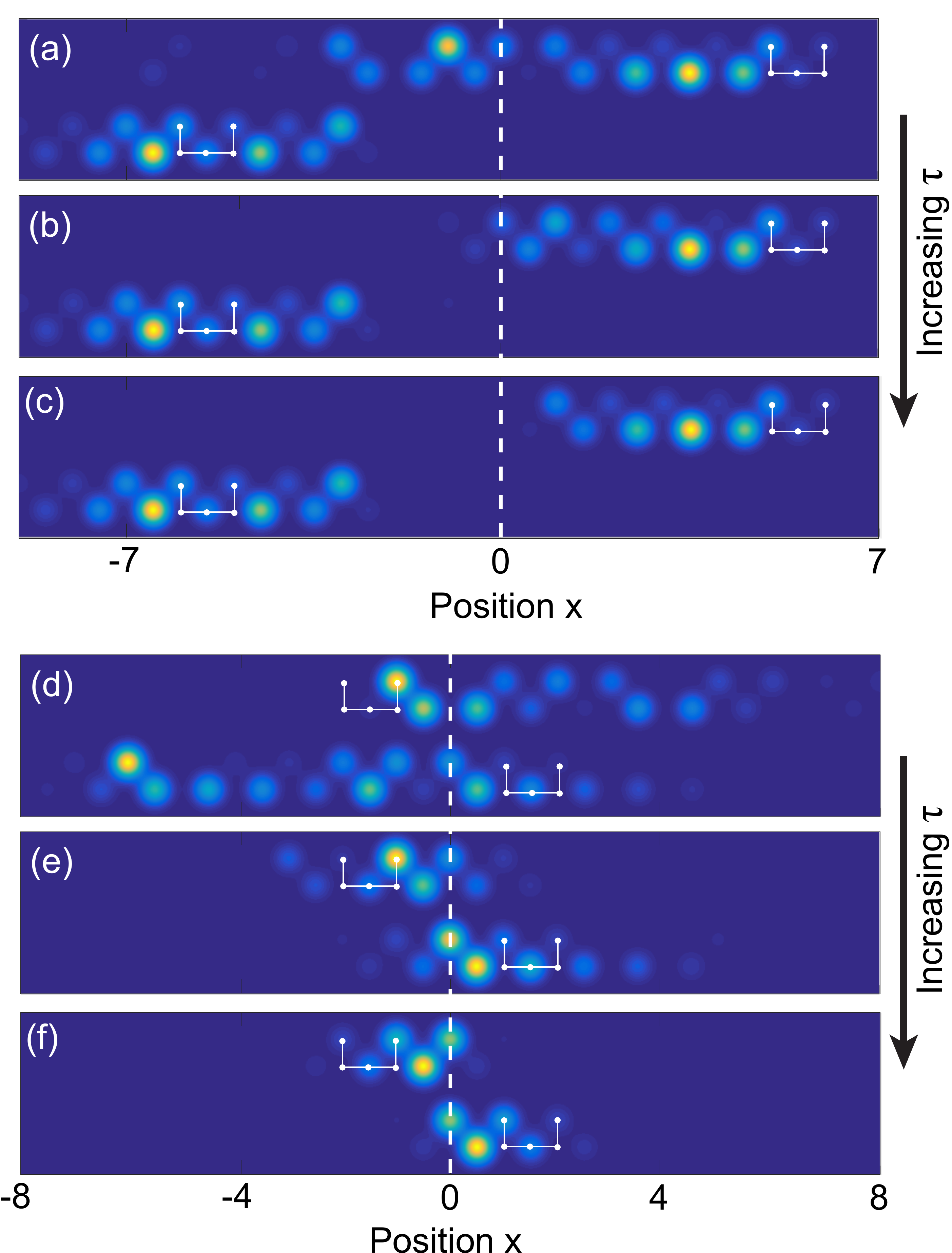}
\caption{(Color online) (a)--(c) Evolution of the dark states in the $N/2-1$ degenerate flat band manifolds from $\tau=5\times10^{-4}$, 0.01, to 0.075. Two typical examples are shown. (d)--(f) Same as in (a)--(c) for the doublet states.
The color scale and legend are the same as in Figs.~\ref{fig:wf}(a) and (b). The amplitude of onsite white noise is $W=0.05$ and $N=20$. }\label{fig:rand}
\end{center}
\end{figure}

Unlike the doublet states, the $\cal PT$-broken dark states in the two degenerate manifolds can be randomly positioned along the lattice, but they are strictly confined on either the gain or the loss half of the lattice.
Due to their degeneracy, any arbitrary superpositions of them are still eigenstates of the system, and one can always find pairs of them such that they are $\cal PT$-symmetric partners of each other. However, the relevant superpositions are determined by the disorder in the system. \cc{Here we consider weak disorders by including a white noise with a uniform distribution in $[-W/2,W/2]~(W\ll G,J)$} around $\omega_0$ on each lattice site. As the lower mode in Figs.~\ref{fig:rand} (a)--(c) shows, if a flat band mode in the two degenerate manifolds is already confined to the loss (or gain) half of the lattice at small $\tau$, increasing $\tau$ has little effect on its spatial profile. However, if at small $\tau$ one of these modes occurs in both the gain and loss regions, it will be forced to take either the gain half or the loss half, as the upper mode in Figs.~\ref{fig:rand} (a)--(c) shows. Here the prevailing effect of $i\tau\bm{H}_1$ is its non-Hermicity, separating the lattice into a gain half and loss half.

In contrast, the prevailing effect of $i\tau\bm{H}_1$ on the doublet states is its point defect mentioned previously. At very small $\tau$, the doublet states are localized by disorder, i.e., they are Anderson localized \cite{Anderson}, and the localization length can be rather long in a particular disorder realization (although on average the localization length is not much longer than that determined by the point defect \cite{PRB15}). As $\tau$ increases, the point defect in $\bm{H}_1$ gradually prevails over the white noise disorder, and the doublet states approach their spatial patterns in the absence of $W$, as can be seen from Figs.~\ref{fig:rand}(e) and \ref{fig:rand}(f) in comparison with Fig.~\ref{fig:wf}. In other words, the $\cal PT$-symmetric relation $\phi_1 = {\cal PT}\phi_2$ of the doublet states
is restored as $\tau$ increases. The same cannot be said about the other flat band modes in general, due to the presence of disorder that breaks the $\cal PT$ symmetry.

\begin{figure}[t]
\begin{center}
\includegraphics[clip,width=\linewidth]{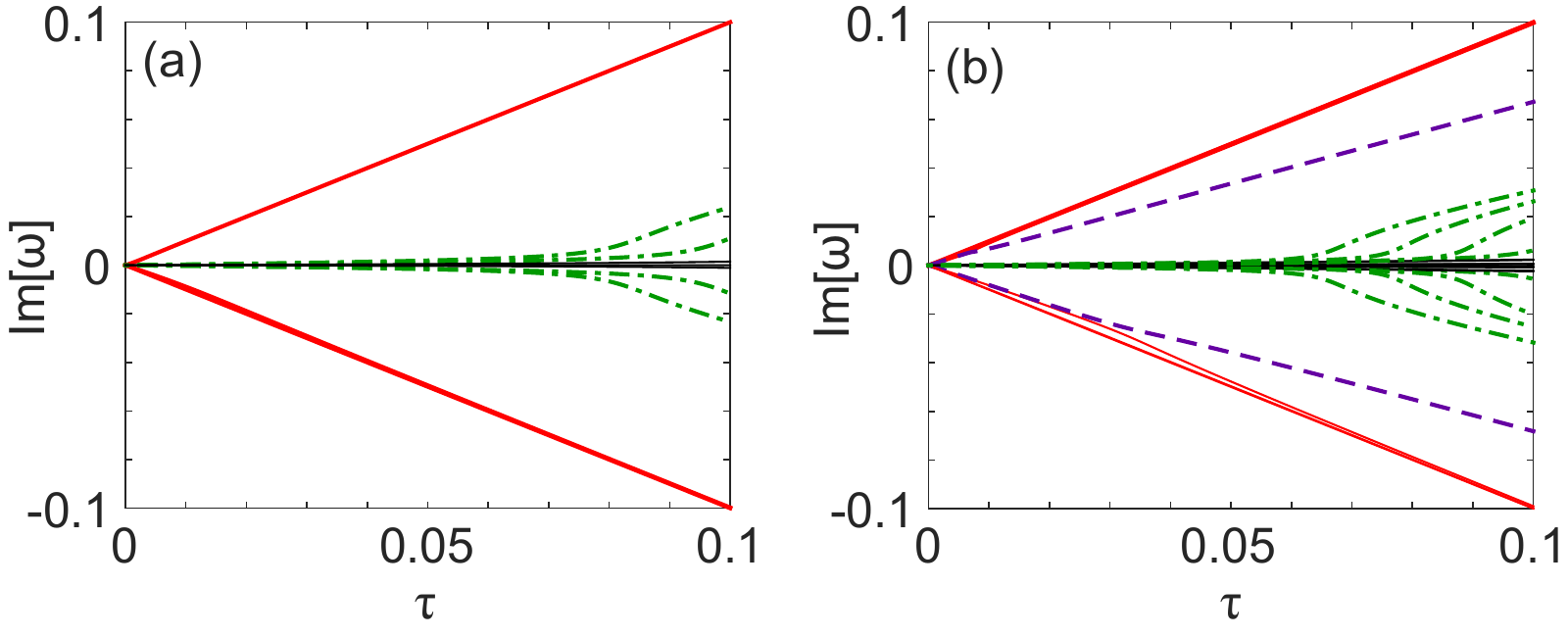}
\caption{(Color online) Same as Fig.~\ref{fig:breaking2} but with disorder. The amplitude of onsite white noise is $W=0.05$. }\label{fig:breaking2_rand}
\end{center}
\end{figure}

Finally, we find that weak disorder has a stronger effect on the dispersive bands than the flat band in terms of $\cal PT$ breaking. As Fig.~\ref{fig:breaking2_rand} shows, the disorder smoothes out the $\cal PT$-transitions of the dispersive band modes in Fig.~\ref{fig:breaking2}, but the thresholdless $\cal PT$ breaking of the flat band modes is largely unaffected. \cc{These contrasting behaviors are due to the different routes to $\cal PT$-symmetry breaking, i.e., whether an exceptional point is involved. On the one hand, an exceptional point is very sensitive to perturbations, and it quickly becomes an avoided crossing in the complex eigenvalue plane when the $\cal PT$-symmetry of the system is lifted by the weak disorder. In fact, each bifurcation of $\im{\omega}$ we have seen in Fig.~\ref{fig:breaking2} from a finite $\tau$ is for two pairs of dispersive band modes as mentioned at the end of Section \ref{sec:II}. Now in the presence of weak disorder, the sensitivity of the exceptional points breaks each of these bifurcations into two, and we see four and eight trajectories of $\im{\omega}$ in Figs.~\ref{fig:breaking2_rand}(a) and (b) (dash-dotted lines) instead of two and four in Figs.~\ref{fig:breaking2}(a) and (b). On the other hand, the thresholdless $\cal PT$-symmetry breaking of the flat band modes does not involve an exceptional point; its role is replaced by the Hermitian degeneracy at $\tau=0$ \cite{PRX14}. In this case there is no singularity, and we have seen that the first-order perturbation theory describes the thresholdless $\cal PT$-breaking well in Section \ref{sec:perturbation}. Now with the weak disorder, its leading (i.e., linear) effect is merely perturbing the real part of the energy eigenvalues, similar to how the $\cal PT$-perturbation varies the imaginary part of the energy eigenvalues. Therefore, these two perturbations are independent processes to the leading order, and the thresholdless $\cal PT$-breaking is largely unchanged with the weak disorder. As the strength of $\cal PT$-breaking increases with $\tau$, the effect of the weak disorder is further suppressed. This can be seen from the following observation.} We find that one flat band mode with $\kappa_-\approx-\tau$ in Fig.~\ref{fig:breaking2_rand}(b) deviates slightly from the other 2 in this branch at small $\tau$ but rejoins the latter as $\tau$ increases. This behavior is due to its spatial profile change caused by the increasing $\cal PT$ perturbation, similar to that of the upper mode in Figs.~\ref{fig:rand}(a)--(c): when this mode has a small portion in the gain half of the lattice at small $\tau$, overall it experiences less loss and hence its negative $\kappa_-$ is larger than $-\tau$. As its spatial profile is squeezed into the loss half of the lattice by the growing $\cal PT$-symmetric perturbation, it experiences more loss than before and its $\kappa_-$ approaches its minimum value $-\tau$. We can also infer that the other 2 modes in this branch are localized already in the loss half when $\tau$ is small, which is verified by inspecting their spatial profiles (not shown).

\section{Conclusion}

We have mentioned that the degeneracy of the flat band modes is completely lifted by a $\cal PT$-symmetric perturbation in general as they undergo thresholdless $\cal PT$ symmetry breaking. An exception takes place in the half-gain-half-loss configuration and its ``V" variant, where two different scenarios of thresholdless $\cal PT$ symmetry breaking occur in the flat band, depending on whether the degeneracy $N$ in the flat band is even or odd. The two degenerate manifolds with $\kappa_\pm=\pm\tau$ always exist and undergo thresholdless $\cal PT$ breaking. While their center positions are random, they are confined to either the gain half or the loss half of the lattice in the absence of disorder. This feature holds even with \cc{weak disorders}, when the $\cal PT$-symmetric perturbation is strong enough. In contrast, the $\cal PT$-symmetric doublet states only exist when $N$ is even, and they display weaker non-Hermicity than the degenerate manifolds. These doublet states are pinned at the symmetry plane with exponential tails in the gain and the loss halves. These tails are the result of a point defect in the $\cal PT$-symmetric perturbation at $x=0$, instead of its half-gain-half-loss nature as previously found \cite{PRA11}. \cc{Weak disorders} may disturb their spatial profiles at small $\cal PT$-symmetric perturbation, but as the latter increases, this feature is restored, together with the $\cal PT$-symmetry relation between the two doublet states.

\acknowledgements
The author thank Hakan T\"ureci, Bo Zhen, and Vadim Oganesyan for helpful discussions. This project is partially supported by NSF under Grant No. DMR-1506987 and by the
Collaborative Incentive Research Grant of City University of
New York, CIRG-802091621.

\appendix
\section*{Appendix: Localization length of the defect states}
In the main text we estimated the localization length of the doublet states by a degenerate perturbation theory. This localization length can also be estimated, for example, using the iterative relation for the values of the wave function on $C$ sites. To derive this iterative relation, we apply the effective Hamiltonian $\bm{H} = \bm{H}_0+\bm{H}'$ to all the lattice sites:
\begin{align}
\Delta_{Aj}A_j &= G B_j,\label{eq:A}\\
\Delta_{Bj}B_j &= G A_j + J(C_j+C_{j-1}),\label{eq:B}\\
\Delta_{Cj}C_j &= J (B_j + B_{j+1}),\label{eq:C}
\end{align}
where $\bm{H}'$ represents any perturbation to the diagonal elements of $\bm{H}_0$ (including the ``V" configuration and the Hermitian defect discussed in Fig.~\ref{fig:wf}) and $Z_j(Z=A,B,C)$ is the wave function on the $j$th site of type $Z$. $\Delta_{Zj}\equiv{\omega}-\omega_{Zj}$, where ${\omega}$ is an eigenvalue of $\bm{H}$ and $\omega_{Zj}$ is the onsite energy of $Z_j$. $\omega_{Zj}$ can have an imaginary part to represent gain or loss if present.

By solving the Eqs.~(\ref{eq:A}) and (\ref{eq:B}), we find
\begin{align}
B_j &= \frac{J(C_j+C_{j-1})}{\Delta_{Bj}-\frac{G^2}{\Delta_{Aj}}}.
\end{align}
By substituting $B_j$ and $B_{j+1}$ in Eq.~(\ref{eq:C}) with this expression, we find the recursive relation for the values of the wave function on $C$ sites:
\begin{align}
&\hspace{6mm}f_j C_j = \chi_{j+1} C_{j+1} + \chi_j C_{j-1},\label{eq:itr4}\\
f_j &= \Delta_{Cj} - \chi_j - \chi_{j+1},\quad
\chi_j = \frac{J^2}{\Delta_{Bj}-\frac{G^2}{\Delta_{Aj}}}.
\end{align}

For the doublet states and its Hermitian counterpart shown in Fig.~\ref{fig:wf}, $\Delta_{Zj}\equiv\Delta_{Z}$ since it is the same for all $Z_j$ lattices on the left (right) side of $x=0$, and $\Delta_{A}=\Delta_{C}\equiv\Delta$. We can then rewrite Eq.~(\ref{eq:itr4}) as
\be
\frac{C_{j+1}}{C_{j}} = \left[\frac{\Delta_{B}\Delta-G^2}{J^2}-2\right] - \frac{C_{j-1}}{C_{j}}. \label{eq:itr5}
\ee
Defining the localization length as $\xi^{-1} \equiv \ln |C_{j}/C_{j+1}| \approx \ln |C_{j-1}/C_{j}|$ for $x>0$ and assuming $|\Delta_B|,|\Delta|\ll G,J$, we find
\be
\xi^{-1}\approx \ln\left(F-\sqrt{F^2-1}\right),\quad F\equiv\frac{G^2}{2J^2}+1 \label{eq:invband}
\ee
using Eq.~(\ref{eq:itr5}), and it gives $\xi\approx1.04$ when $G=J$. This result holds whenever there is a defect state in the Lieb lattice, and it applies to the point defect in both $\bm{H}_1$ of the ``V" configuration and the Hermitian defect we discussed in Fig.~\ref{fig:wf}. The difference of these two cases lies in the values of $\Delta_B,\Delta$, which are not important for $\xi$ as long as $|\Delta_B|,|\Delta|\ll G,J$. Finally, we note that by defining a complex wave vector $k\equiv -i\xi^{-1}$ for $x>0$, Eq.~(\ref{eq:invband}) can be interpreted as solving the band structure of the dispersive bands inversely \cite{band}: instead of finding the energy of a band at a given wave vector, one can find the wave vector at a given energy; if this energy is in the band gap, one finds that $k$ has to be purely imaginary and its absolute value is given by Eq.~(\ref{eq:invband}) when $|\Delta_B|,|\Delta|\ll G,J$.


\end{document}